\documentclass[aps,prl,twocolumn,longbibliography,superscriptaddress]{revtex4-2}
\usepackage{natbib,graphicx,dcolumn,color,lipsum,hyperref}
\usepackage{amsmath,amssymb,verbatim,moreverb,bm,framed} 
\usepackage{siunitx} 
\usepackage{ulem} 

\def\be{\begin{equation}}
\def\ee{\end{equation}}
\def\ber{\begin{eqnarray}}
\def\eer{\end{eqnarray}}

\def\bsigma{\mbox{\boldmath $\sigma$}}
\def\br{{\bf r}}

\def\bp{{\bf p}}

\def\bn{{\bf n}}
\def\bs{{\bf s}}

\newcommand\commentout[1]{}

\def\xc{{\rm xc}}
\def\so{{\rm so}}

\def\be{\begin{equation}}
\def\ee{\end{equation}}
\def\ber{\begin{eqnarray}}
\def\eer{\end{eqnarray}}

\newcommand{\ie}{{\it i.e.~}} 	
\newcommand{\eg}{{\it e.g.~}} 	

\def\xc{{\mathrm xc}}
\def\so{{\mathrm so}}
\def\EY{{\mathrm EY}}

\definecolor{greenS}{rgb}{0.00, 0.6, 0.00}
\definecolor{orangeS}{rgb}{0.6, 0.1, 0.1}

\begin{document}
 \title{Spin charge conversion in Rashba split ferromagnetic interfaces}

 

\author{Olivier Rousseau}
\affiliation{SPEC, CEA, CNRS, Universit\'e Paris-Saclay, 91191 Gif-sur-Yvette, France}
\author{Cosimo Gorini}
\affiliation{Institut f\"ur Theoretische Physik, Universit\"at Regensburg}
\affiliation{SPEC, CEA, CNRS, Universit\'e Paris-Saclay, 91191 Gif-sur-Yvette, France}
\author{Fatima Ibrahim}
\affiliation{SPINTEC, CEA, CNRS, Univ. Grenoble Alpes, 38054 Grenoble, France}
\author{Jean-Yves Chauleau}
\affiliation{SPEC, CEA, CNRS, Universit\'e Paris-Saclay, 91191 Gif-sur-Yvette, France}
\author{Aurélie Solignac}
\affiliation{SPEC, CEA, CNRS, Universit\'e Paris-Saclay, 91191 Gif-sur-Yvette, France}
\author{Ali Hallal}
\affiliation{SPINTEC, CEA, CNRS, Univ. Grenoble Alpes, 38054 Grenoble, France}
\author{Sebastian T\"{o}lle}
\affiliation{Institut f\"ur Theoretische Physik, Universit\"at Augsburg}
\author{Mair Chshiev}
\affiliation{SPINTEC, CEA, CNRS, Univ. Grenoble Alpes, 38054 Grenoble, France}
\author{Michel Viret \thanks{corresponding author: \texttt{michel.viretr@cea.fr}}}
\affiliation{SPEC, CEA, CNRS, Universit\'e Paris-Saclay, 91191 Gif-sur-Yvette, France}

 \date{\today}

\begin{abstract}
We show here theoretically and experimentally that a Rashba-split electron state inside a ferromagnet can efficiently convert a dynamical spin accumulation into an electrical voltage.
The effect is understood to stem from the Rashba splitting but with a symmetry linked to the magnetization direction.
It is experimentally measured by spin pumping in a CoFeB$|$MgO structure where it is found to be as efficient as the inverse spin Hall effect at play when Pt replaces MgO, with the extra advantage of not affecting the damping in the ferromagnet.
\end{abstract}
\maketitle


Classical spintronics relies on the generation and manipulation of spin-polarized electrical currents in magnetic conductors.
Enormous progress have been achieved in the past 30 years to harvest the spin effects and use them in specific devices.
In most cases, it turns out that the electrical part of these currents is detrimental for applications as it brings undesired effects like Joule heating.
Recently, it has been realized that pure spin currents can be generated and propagated in non-magnetic materials or even electrical insulators.  
Here the key player is the spin-orbit coupling (SOC) interaction, at the heart of a new field called spin-orbitronics \cite{Manchon2019}.
Spin-to-charge interconversion effects are therefore a subject of intensive theoretical and experimental investigation.
Possibly the best known of these is the Spin Hall Effect (SHE) first introduced in the seventies \cite{DyakonovPerel1971a, DyakonovPerel1971}.
It relies on a preferential directional scattering of electrons of different spins by SOC or on particular band structures referred as extrinsic and intrinsic SHE mechanisms, respectively.
It results in the generation of a transverse (pure) spin current when a charge current flows in a large SOC material \cite{Hirsch1999} like, for instance, the heavy metal Pt.
The reciprocal to SHE is the inverse Spin Hall Effect (ISHE), which has been widely used recently to detect a spin current as the SOC interaction converts it into a (transverse) charge current \cite{Kimura2007}.
The effect is particularly interesting to sensitively measure magnetization dynamics, as it is accompanied by spin current emission peaking at the ferromagnetic resonance\cite{Saitoh2006}.
However, as far as this type of measurement is targeted, a conducting SOC layer in contact with the probed ferromagnet usually induces a very significant extra source of damping related to the spin pumping effect\cite{Mizukami2001, TserkovnyakPRL2002}. Hence, the ISHE measurement is rather intrusive.
For this purpose, another effect is better suited. It relies on the charge current generation from a non-equilibrium spin population injected into a spin-orbit coupled system,
and goes under the name of spin-galvanic effect (SGE)\cite{ganichev2003}. It is often also called inverse Edelstein effect (IEE) if it takes place in a two-dimensional (2D) Rashba state, as will be the case here\footnote{For details, see \eg Refs.~\cite{ganichev2016,ivchenko2017,gorini2017}}.
It has been demonstrated in systems like Ag$|$Bi \cite{RojasSanchez2013, Sangiao2015}, Fe$|$GaAs \cite{Chen2016} or even LAO$|$STO \cite{Lesne2016, Chauleau2016}.
It can be qualitatively understood recalling that, in the presence of SOC, which alone does not break time-reversal symmetry, an electronic ensemble can acquire a non-equilibrium global spin polarization if drifting -- that is, if a charge current is present.  In turn, pumping spins will generate a charge current arising from a balance between at least three phenomena: (i) spin relaxation, limiting the spin pumping efficiency; (ii) SOC, transferring the spin information to momentum; (iii) momentum relaxation, limiting drift motion. The detailed microscopic equilibrium can however be subtle \cite{Shen2014,ganichev2016,toelle2017,ivchenko2017,gorini2017}.
The IEE presents some advantages because a 2D gas is generally not too invasive and it is more efficient at generating
 high voltages than a Pt layer as its ``internal resistance" is higher. Another important parameter for the conversion efficiency is the effective interface spin conductance, $G_{\uparrow\downarrow}$,
describing the interface transparency to the spin accumulation \cite{TserkovnyakPRL2002}. Rather high for a metal$|$metal contact (Co$|$Pt for example), it is generally a factor of 10 lower for a metal$|$insulator interface \cite{Zhang2015, Hahn2013} inducing a dramatic loss of efficiency for tunnelling systems like LAO$|$STO.

An ideal probe to measure magnetization dynamics would thus be one where spin-to-charge conversion neither affects the magnetic properties nor decreases the device resistance,
 while providing an optimum spin transmission to the convertor. It seems therefore best if a high resistance Rashba-split state could be located \textbf{inside} the magnetic layer.
 In order to find the right system, theoretical studies on electric field control of magnetic anisotropy at FM$|$insulator interfaces provide some key insights, as they reveal the existence of rather large electric dipoles ($\approx2.2{\rm V/nm}$) at the Fe$|$MgO interface \cite{Ibrahim2016}.
These produce Rashba-split interfacial states \cite{Dieny2017} mediating the Dzyaloshinskii-Moryia interaction \cite{Yang2018b}. The band-splitting is indeed significant near the Fermi level as it produces a Rashba parameter of the order of 224 meV.\si{\angstrom}. These findings trigger our interest in investigating CoFe$|$MgO interfaces hosting a Rashba state and harvest it for spin-to-charge conversion. In this paper, we thus calculate this interface band structure and lay the theoretical basis for the spin-to-charge conversion properties of the resulting ferromagnetic Rashba interface. We finish by the experimental demonstration of its efficiency.

\begin{figure}[h!]
\includegraphics[width=\linewidth]{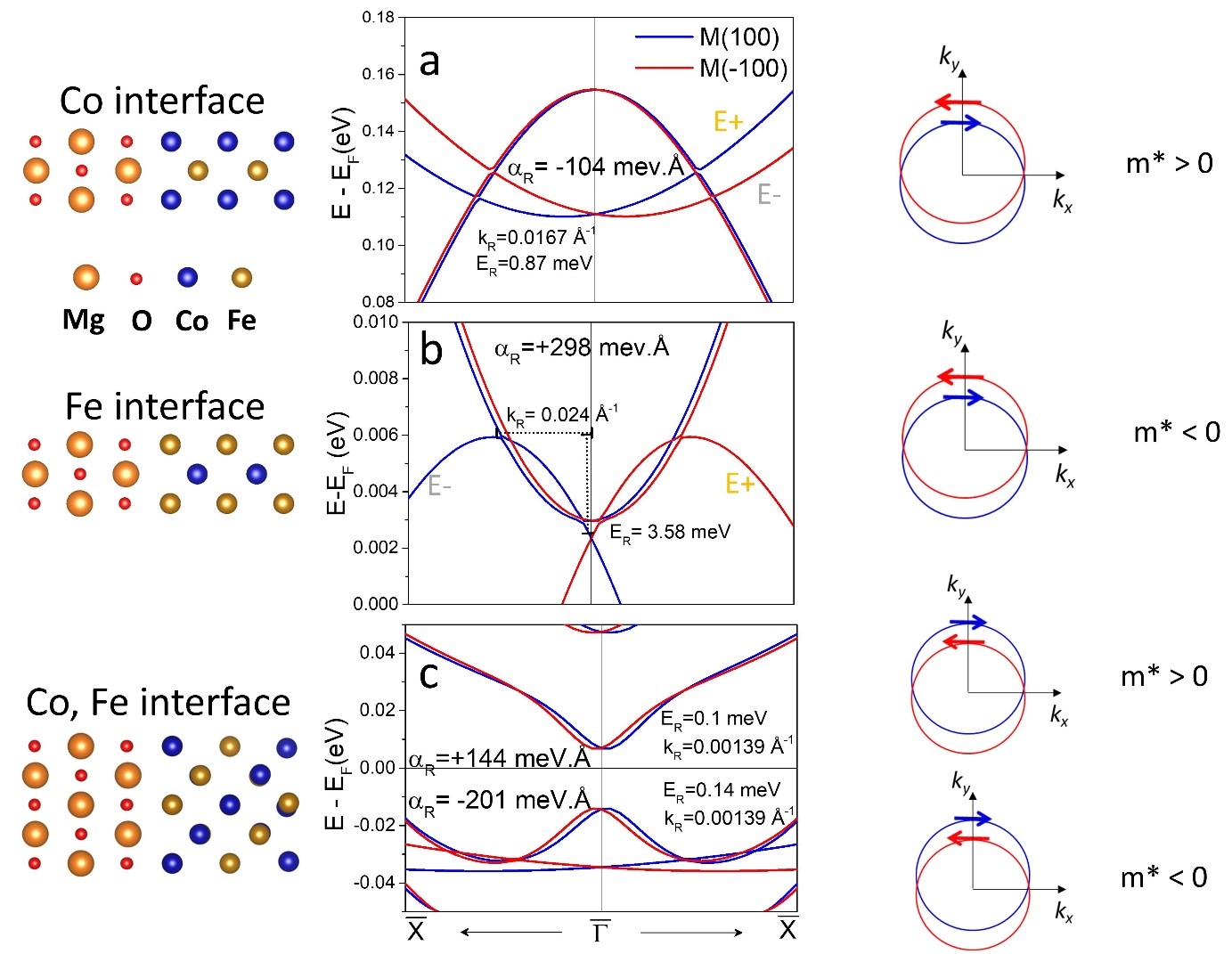}
\caption{The Rashba splittings $\alpha_R$ calculated from the band structures ($E vs k$) of (a) Co-, (b) Fe-, and (c) Co,Fe-terminated CoFe/MgO interfaces are displayed in the inset of the graphs. The right column represents Fermi surfaces corresponding to opposite magnetization orientations for each interface with the associated sign of the effective mass $m^*$. 
}.
\label{FigFatima}
\end{figure}

We performed systematic first-principles calculations of the electronic band structure including spin-orbit interaction at the CoFe$|$MgO (001) interface considering different terminations. 
Our calculations are based on the projector-augmented wave (PAW) method \cite{Blochl1994} as implemented in the VASP package \cite{Kresse1993, Kresse1996} using the generalized gradient approximation \cite{Perdew1996} and including spin-orbit coupling.
The calculated supercell comprises five monolayers of bcc-CoFe(001) on top of three monolayers of MgO (001) followed by a sufficiently large vacuum layer of 25\si{\angstrom}.
The in-plane lattice parameter is fixed to that of CoFe,\, \ie 2.83\si{\angstrom}. Three different interfacial terminations are considered: Co, Fe, and Co-Fe. A kinetic energy cutoff of 550 eV is used for the plane-wave basis set and 25×25×1 K-point mesh to sample the first Brillouin zone. A larger supercell (2×2) is used to model the Co-Fe terminated interface with a 15×15×1 K-point mesh.
Calculations are performed in three steps. First, we let the structures relax until the forces become smaller than 1 meV/\si{\angstrom}.
 Next, the Kohn-Sham equation is solved without SOC to find out the system's ground state charge density. Finally, SOC is included and the band structure is calculated as a function of the magnetization orientation pointing along (100) and (-100).
By superimposing these band structures with opposite magnetization directions, the Rashba effect can be observed as the free-electron like surface states are split in energy by the potential gradient perpendicular to the surface. This leads to the following dispersion relation: 
\be
\label{Disp}
E_{\pm} (k)=E_0+\frac{\hbar^2 k^2}{2m^*} \pm |\alpha_R| |k| .
\ee
where $ E_0$ is the energy at which the band crosses $k=0$ and $m^*$ is the effective electron mass, whose sign determines whether $E_\pm$ is the upper or lower branch.
The Rashba coefficient is defined as $ \alpha_R = 2E_R/k_R $, with $E_R$ the Rashba splitting at $k=k_R$ (the k vector of the  splitting).
 For magnetic systems, the spin polarization of electronic states results from the exchange coupling, so the Rashba SOC scales down from the maximum value when $k\parallel M$ to zero when $k\perp M$. The sign of the Rashba coefficient is determined by the spin polarization and the sign of $m^*$ \cite{Henk2004,Bentman2011,Barnes2015}. For each termination, $\alpha_R$ is estimated from the band splittings closest to the Fermi energy, calculated with magnetization axis M pointing along (100) and (-100) directions (blue and red lines in Fig.~\ref{FigFatima}). 

The spin orientation of the Rashba branches are represented by red and blue arrows for each interface in Fig.~\ref{FigFatima}. 
For Co and Fe terminations, we find the presence of Rashba-split bands with positive and negative effective masses $m^*$, respectively, located slightly above the Fermi level.
The corresponding Rashba coefficients are found to be $\alpha_R = -104 (+298)$meV.\si{\angstrom} for Co (Fe) terminations, respectively. 
For the case of the Co-Fe terminated interface, two Rashba-split bands are present giving rise to $\alpha_R = -201 (+144)$meV.\si{\angstrom} below (above) the Fermi level.
Such values are comparable to those reported at Co$|$MgO(111) interfaces, and larger than that obtained for other Rashba interfaces with magnetic layers such as Co$|$graphene \cite{Yang2018b}.  
We point out that the shown Rashba-split bands are chosen to be the closest to the Fermi energy and originate from minority spin down bands for Co- and Fe- terminated interfaces and majority spin up bands in the case of Co-Fe termination.

\begin{figure}[h!]
 \includegraphics[width=.9\linewidth]{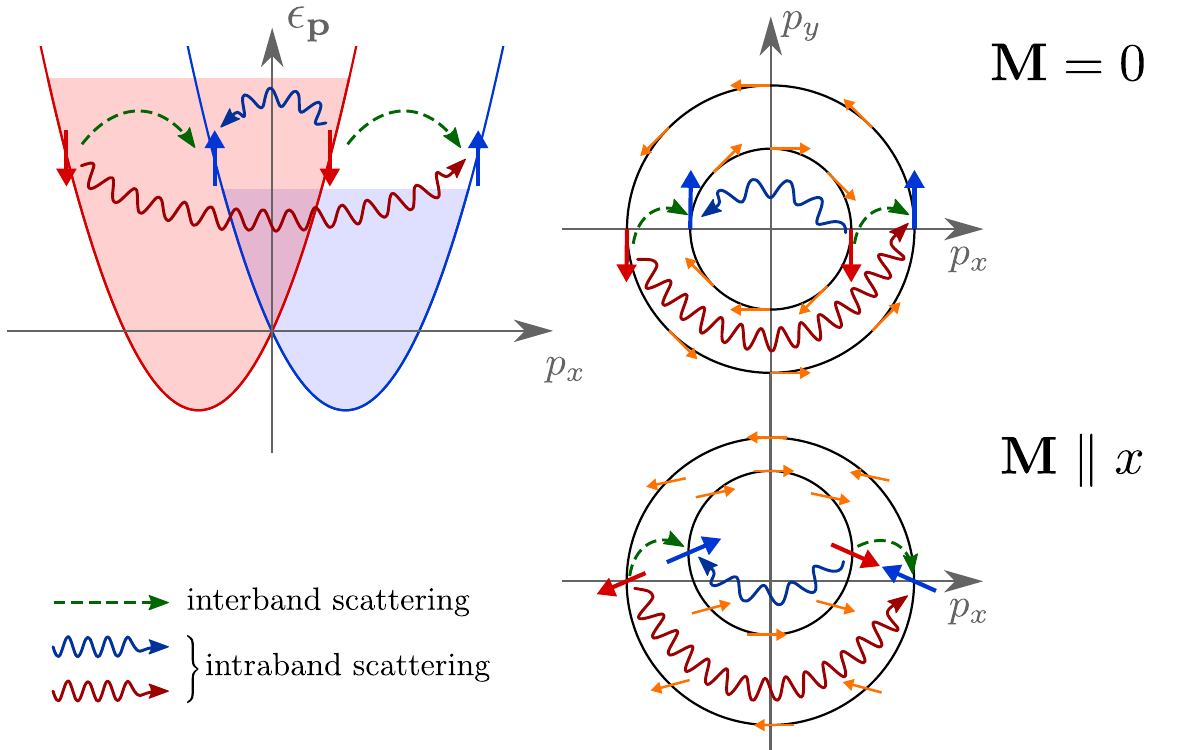}
\caption{Sketch of current generation from an injected non-equilibrium spin polarization $\delta {\bf s}$ 
in Rashba spin-orbit-split subbands (unlike Fig. 1, both bands are represented), following Refs.~[\onlinecite{ganichev2016,ivchenko2017}].  
Spin-flip events due to Elliott-Yafet happen at a rate $\hbar/\tau_\EY$
and redistribute electrons so that steady-state is eventually reached, $\delta\bs \to \langle\delta\bs\rangle$.
Interband scattering (green lines) compensate each other, while intraband (red and blue lines) do not,
by an amount $(N^+-N^-)/N_0 \sim \Delta_\so/\epsilon_F$.  
A current is established on the scale of the momentum relaxation time $\tau$,
meaning that only a fraction $\tau/\tau_\EY$ of the spin-flip processes contribute to its generation,
yielding Eq.~\eqref{estimate}.  In the presence of strong exchange $\Delta_\xc \gg \Delta_\so$,
the same arguments apply, as the number of Fermi surface electrons yielding the Rashba
spin-momentum texture remain $\sim\Delta_\so/\epsilon_F$.  However only the component
of the non-equilibrium spin polarization parallel to the strong exchange field,
$\langle\delta{\bf s}\rangle\parallel\bn$, can generate a current.  Fast precession
$\Delta_\xc\gg\hbar/\tau$ results in $\langle\delta\bs\rangle\perp\bn\approx 0$.}
\label{fig_handwaving}
\end{figure}

It is now necessary to understand how the spin-charge conversion operates in such 
strongly magnetic interface states, and how it compares to the ``conventional'' effect of a pure non-magnetic Rashba state.
This is done within the theoretical framework of Ref.~[\onlinecite{toelle2017}],
but it is instructive to first follow the reasoning established for non-magnetic semiconducting quantum wells
\cite{ganichev2003,ganichev2016,ivchenko2017}, and later suitably generalize it.     
We start from the low-energy effective model
\be
\label{H1}
H = H_{\rm R} + H_\xc + H_{\rm imp}.
\ee
Here $H_{\rm R}=\hbar^2k^2/2m^* + \alpha_R \left(k_y\sigma^x-k_x\sigma^y\right)$ describes a 2D
Rashba system, yielding a Fermi surface spin splitting $\Delta_\so = 2\alpha_R k_F$.
$H_\xc=(\Delta_\xc/2){\bf n}(\br,t)\cdot\bsigma$ is an $s$-$d$ exchange coupling with an arbitrarily pointing
homogeneous magnetization ($|{\bf n}|=1$).  
Here upper (lower) indices denote spin (real space) components, and summation over repeated indices is implied.
Finally, $H_{\rm imp}=V(\br)+(\lambda^2/4\hbar)\bsigma\times\nabla V(\br)\cdot\bp$ models
the inteaction with a random disorder potential $V(\br)$, taken of standard white-noise (short range) form, with $\lambda$ the effective Compton wavelength).
Such a potential directly causes momentum and Elliott-Yafet spin flip relaxation, respectively with rates
$\hbar/\tau, \hbar/\tau_{\rm EY}$ \cite{raimondi2009,raimondi2012}, and, in conjuction with $H_{\rm R}$,
Dyakonov-Perel spin relaxation $\hbar/\tau_{\rm DP}$.
The standard picture of spin-to-charge conversion relies on drift-diffusion 
arguments \cite{ganichev2016,ivchenko2017,gorini2017}, which require $\Delta_\so, \Delta_\xc \ll \hbar/\tau$.
Crucially, this picture breaks down in a strong magnetic state defined by
\be
\Delta_\xc \gg \Delta_\so, \hbar/\tau.
\ee
Nonetheless, consider first a diffusive non-magnetic state, $\Delta_\xc=0$, 
in which a DC non-equilibrium spin polarization
$\langle\delta s^{x,y}\rangle$ is generated by some means, where $\langle\dots\rangle$ stands for time average.
The Rashba spin texture connects a positive $s^x$ with
a negative $k_y$ and a positive $s^y$ with a positive $k_x$ for the majority band $N^+=N_0+\delta N^+$, 
the opposite holding true for the minority band $N^-=N_0+\delta N^-$, see Fig.~\ref{fig_handwaving}.  
Here $N_0$ is the density of states per spin and volume of a 2DEG without Rashba coupling,
while $\delta N^{\pm}/N_0 \sim \mp \Delta_\so/\epsilon_F \ll 1$.  
The electrons building up the non-equilibrium steady-state $\langle\delta s^{x,y}\rangle$ 
are thus anisotropically distributed around the two spin-orbit split circles, 
with a skewedness $\sim\Delta_\so/\epsilon_F$.  A DC current ($J$) is established on the scale
of the momentum relaxation time $\tau$, so that of these skewedly-distributed electrons,
a fraction $\tau/\tau_\EY$ contributes to its generation. This leads to the estimate
\be
\label{estimate}
J_{x,y} \sim -e v_F \; \frac{\tau}{\tau_\EY} \frac{\Delta_\so}{\epsilon_F} \; \langle \delta s^{y,x} \rangle
= \frac{-e\alpha_R\tau}{\tau_\EY} \; \langle \delta s^{y,x} \rangle,
\ee
as experimentally observed \eg in Ref.~[\onlinecite{ganichev2003}].

Now consider instead spin pumping at (microwave) frequency $\omega$ into a strong ferromagnet without any Rashba coupling,
$H_{\rm R}=0$. The general result \cite{tserkovnyak2008}
$\delta {\bf s} = (\hbar N_0/2)[\bn \times \dot{\bn}-(\hbar/\Delta_\xc\tau_{\rm EY})\dot{\bn}]$  
yields for our geometry 
$\langle\delta s^x\rangle \sim \omega N_0 \cos\phi\sin^2\theta,\;
\langle\delta s^y\rangle \sim \omega N_0 \sin\phi\sin^2\theta,\;
\langle\delta s^z\rangle \sim 0$.
Clearly, only a DC non-equilibrium spin polarisation along the direction
of the strong exchange field is possible -- orthogonal contributions relax too fast to establish themselves.
The electrons making up $\delta {\bf s}$ are {\it e.g.} electrons from the minority band $N^\downarrow$ whose spin is flipped
by the driving field, so that they join the majority band $N^\uparrow$.
 Without Rashba spin-orbit coupling such electrons depopulate/populate $N^\downarrow/N^\uparrow$ isotropically in momentum space around the Fermi circles, resulting in no DC charge current.

Let us now consider Rashba splitting in this strongly magnetic state and estimate the SGE/IREE driven by spin pumping,
$H_{\rm R}\neq0, \Delta_\xc \gg \Delta_\so, \hbar/\tau$.
Exchange dominates, so to leading order $\langle\delta {\bf s}\rangle$ is still given by its $H_{\rm R}=0$
expression \cite{tserkovnyak2008}.
However the electrons making up $\langle\delta {\bf s}\rangle$ are distributed anisotropically in momentum
space, as both minority and majority bands are slightly distorted by $H_{\bf R}$, \ie
$N^{\uparrow\downarrow} \to N^{\uparrow\downarrow} + \delta N^\pm$. 
Thus the same arguments used in the $\Delta_\xc=0$ case, see Fig.~\ref{fig_handwaving},
can now be exploited for the fractions $\delta N^\pm$ of the overall electronic population 
generating the skewed distribution, leading again to Eq.~\eqref{estimate}. 
One concludes
\ber
J_x &\sim& -\frac{e\alpha_R\tau}{\tau_{\rm EY}} \langle \delta s^y\rangle
=
-\frac{e\alpha_R\tau}{\tau_\EY} \omega \sin\phi \sin^2\theta,
\\
J_y &\sim& -\frac{e\alpha_R\tau}{\tau_{\rm EY}}\langle \delta s^x\rangle
= 
-\frac{e\alpha_R\tau}{\tau_\EY} \omega \cos\phi \sin^2\theta.
\eer
Up to numerical prefactors of order 1, these estimates are confirmed by microscopic calculations \cite{toelle2017}
obtained within an $SU(2)$ kinetic formulation \cite{gorini2010,raimondi2012,gorini2017}.  In a more general
vector form one has
\be
\label{SGE_vectorial}
{\bf j}^{\rm SGE} = 2\,e\alpha_R\omega N_0\,\frac{\tau}{\tau_\EY}
\frac{\langle{\bf e}_z\times(\bn\times\dot{\bn})\rangle}{\omega}.
\ee
Note that there exists a second spin-to-charge channel, in which the conversion is mediated by the inverse spin
Hall effect in the 2D plane \footnote{This process is generated by an $SU(2)$ diffusive contribution to the
spin current \cite{gorini2010,raimondi2012}, and is related to the ``\textit{relaxational contribution}'' discussed 
in the review [\onlinecite{ivchenko2017}].}.  
This mechanism relies crucially on precession around the spin-orbit field and might be dominant in a non-magnetic 2DEG \cite{Shen2014}.
It becomes however negligible in the presence of strong exchange \cite{toelle2017}, since the slow precession around
the spin-orbit field $(\Delta_\so\ll\hbar/\tau)$ is disrupted by the fast one around the exchange field 
$(\Delta_\xc\gg\hbar/\tau)$.  Explictly one obtains
\ber
{\bf j}^{{\rm ISHE}\to{\rm SGE}} \sim \left(\frac{\Delta_\so}{\Delta_\xc}\right)^2 {\bf j}^{\rm SGE} \ll {\bf j}^{\rm SGE}.
\eer

It is important to note here that the non-equilibrium spin population at the heart of the effect in the magnetic Rashba case is also mainly responsible for the intrinsic damping of the magnetic layer. No significant extra damping is thus added in this process.

This spin to charge conversion due to a ferromagnetic Rashba interfacial state should thus be measurable in Fe$|$MgO systems. Experimentally, because of its superior dynamical properties, ${\rm Co}_{0.4}{\rm Fe}_{0.4}{\rm B}_{0.2}$ is used as the magnetic material.
Layers of various thicknesses ranging from 3 to 20 nm were deposited by sputtering (Singulus system) on float glass substrates
 capped with 3 nm of Pt or 5 nm of MgO in order to provide two classes of samples relying on different conversion mechanisms.
 Spin injection was achieved by spin-pumping \cite{Saitoh2006}: the magnetic layer is excited by a microwave magnetic field at a fixed frequency.
 Sweeping a DC magnetic field permits to cross the ferromagnetic resonance which is detected by a diverging microwave absorption.
 Noteworthy the applied DC magnetic field is slowly modulated which enables a lock-in detection of the derivative of the dynamic magnetic susceptibility with respect to the applied magnetic field \cite{Chauleau2016}. 

\begin{figure}[h!]
\includegraphics[width=\linewidth]{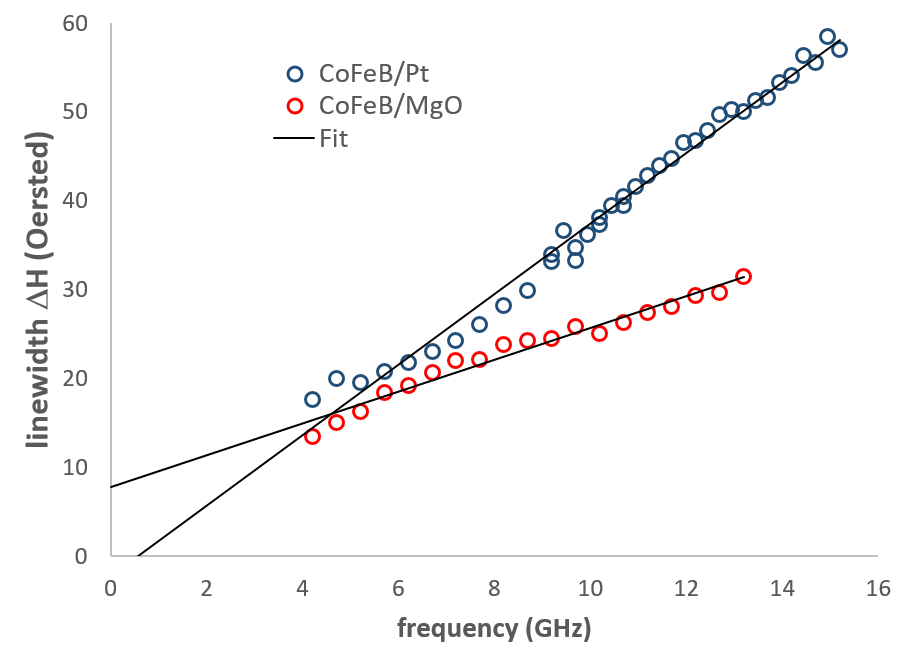}
\caption{Extraction of the damping (from the slopes of $\Delta H$ vs frequency) in two different CoFeB layers capped respectively with 3 nm of Pt (red dots) and 5 nm of MgO (blue dots). The former induces an extra damping of the order of $\Delta\alpha=5.10^{-3}$.}.
\label{FigFMR}
\end{figure}

 Using Kittel's theory, one can quantify the magnetization and damping that is responsible for dissipating angular momentum.
 Our bare CoFeB layers have typical intrinsic damping constants around $\alpha=5.10^{-3}$, unaltered by an adjacent MgO layer, but a 3 nm thick Pt layer brings it to above $10^{-2}$ as shown in Fig.~\ref{FigFMR}.

\begin{figure}[h!]
 \includegraphics[width=\linewidth]{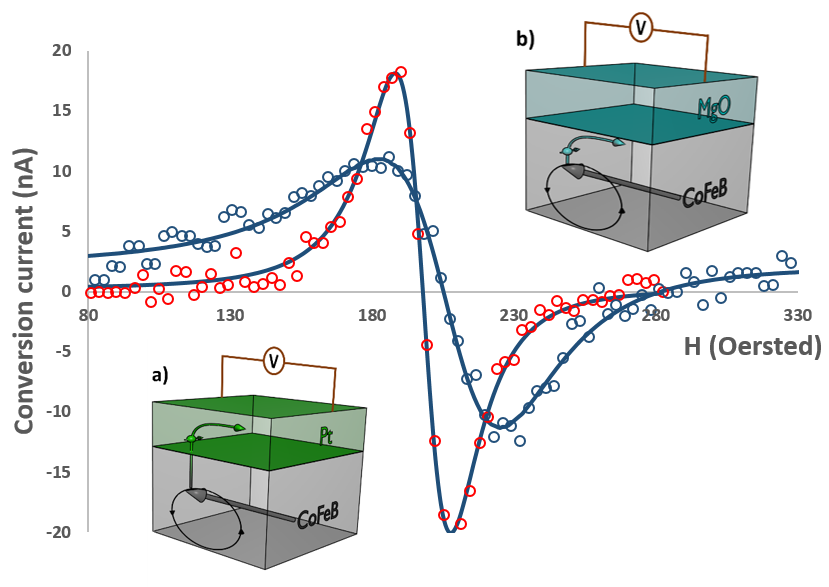}
\caption{Comparison of the (derivative of) conversion charge current between CoFeB(5nm)$|$Pt(3nm) in blue and CoFeB(5nm)$|$MgO(5nm) in red (with respective resistances of $138 \Omega$ and $245 \Omega$), at 4.4 GHz. The measured voltage is perpendicular to the magnetization direction as shown in the sketches a and b. The Rashba state at the CoFeB$|$MgO interface is a very efficient spin-to-charge convertor as even when considering the scaling with the (inverse square of the) linewidth, it is only about 20\% less than the ISHE in Pt.}
\label{CompFig}
\end{figure}

 In parallel, we also measure the conversion voltage for both Pt and MgO capped CoFeB layers. The samples with dimensions of $8mmx10mm$ are placed in our broad band RF cell with contacts near the edges of their 10mm length. An extra varying field of 3 Oe at a frequency of 23 Hz allows for a lock-in measurement of the field derivative of the transverse voltage generated at FMR resonance.
 Interestingly, the measured voltages are larger for the CoFeB$|$MgO systems, indicating a very efficient spin-to-charge conversion in this system.
Even converted into a current (the real physical quantity), the effect still exceeds that in CoFeB/Pt as can be seen in Fig. \ref{CompFig} where it is shown for two representative samples. Taking into account the difference in resonance width, and recalling that the amplitude of the Lorentzian derivative scales with $\frac{1}{\Delta H^2}$, 
one concludes that Pt is only 20\% more efficient than the MgO interface. 
We also verified with the full angular dependence of the signal with in-plane field angle, that the symmetry of the effect as is consistent with spin/charge conversion, where AMR rectification is negligible (not shown here).
The spin pumping efficiency into the Pt depends on several parameters but the crucial ones are the effective interface transparency $g_{\uparrow\downarrow}^{eff}$ and the Spin Hall angle $\theta_{SH}$. For low enough frequency ($\omega \ll \gamma M_s$), one gets for the injected spin current:

\be
j_{s} = \frac{\hbar}{4\pi} g_{\uparrow\downarrow}^{eff} \frac{\gamma h_{rf}^2}{4\pi M_s \alpha^2},  
\quad g_{\uparrow\downarrow}^{eff}=\frac{4\pi M_s t_F}{g\mu_B}\Delta\alpha.
\ee

where $ t_F$ is the ferromagnetic layer's thickness, $ M_s$ its magnetization and $\Delta\alpha$ the extra damping generated by the Pt layer. The conversion voltage then reads:

\be
V_{ISH}=\frac{e}{2\pi} \frac{L}{\sigma_N t_N + \sigma_F t_F} J_s \theta_{SH} \,\tanh\left(\frac{t_N}{2\lambda_{sd}}\right).
\ee

with $\sigma$ and $t$ the conductances and thicknesses of the normal metal and the ferromagnet and $\lambda_{sd}$ the spin diffusion length in Pt.
For our samples, we find $g_{\uparrow\downarrow}^{eff}\approx 3.10^{19}$ and $\theta_{SH}\approx 0.1$ taking $\lambda_{sd}=3\,{\rm nm}$.

For the Inverse Edelstein effect, the Rashba splitting replaces the spin Hall angle as the relevant quantity. It is predicted that a Rashba splitting of $0.3$ eV.\si{\angstrom} should be equivalent to the spin Hall angle of Pt \cite{toelle2017}.
Considering that in our samples the dynamical angular momentum dissipation is mainly through relaxation by the electron system, \ie $\alpha$ is mainly due to electronic effects, the above model predicts a spin-to-charge conversion slightly smaller for CoFeB/MgO compared to CoFeB/Pt, in excellent agreement with the measurements. 
It is to be noticed here that because the active area is the interface between CoFeB and MgO, it is very sensitive to the exact deposition conditions and it may not exist if the interface is intermixed or badly oxidized. As a matter of fact, some existing reports could not measure spin/charge conversion in Fe deposited on MgO \cite{RojasGe2013,oyarzun2016}, and interestingly, in our samples the effect only appears when MgO is deposited on CoFeB but not for the reversed structure.

To conclude, we have shown here that the Rashba-split state existing at the interface between CoFeB and MgO is able to efficently convert the non equilibrium spin population generated by FMR spin pumping into a charge current.
The effect is understood as an inverse Edelstein effect (or spin galvanic) stemming from the ferromagnetic Rashba-split state. It is as efficient as that using the ISHE in a Pt layer with the added value of not deteriorating the resonance properties of the ferromagnet.

\vspace{5mm}

\begin{acknowledgments}
We would like to acknowledge funding from the French National Research Agency (ANR) through the SANTA project (18-CE24-0018-01) and the project SPICY from the Labex NanoSaclay “Investissements d’Avenir” program (reference: ANR-10-LABX-0035).
\end{acknowledgments}

\bibliographystyle{apsrev4-1}

%

\end{document}